\documentclass[twocolumn,12pt]{article}
\usepackage{times}

\usepackage[english]{babel}
\makeatletter
\usepackage{amsmath}
\usepackage{fancyhdr}
\usepackage{cuted}
\makeatother
\usepackage{afterpage}
\usepackage[letterpaper,top=1cm,bottom=2.5cm,left=2cm,right=2cm,marginparwidth=1.75cm]{geometry}
\usepackage{paracol}  
\usepackage{gensymb}
\usepackage[utf8]{inputenc}
\usepackage{graphicx}

\usepackage[version=4]{mhchem}
\usepackage{amsmath}
\usepackage{graphicx}
\usepackage[colorlinks=true, allcolors=blue]{hyperref}
\title{Tunable Ultra-Strong Magnon-Magnon Coupling Approaching the Deep-Strong Regime in a van der Waals Antiferromagnet}

\author{
  C. W. F. Freeman$^{1,2,*}$,
  H. Youel$^1$,
  A.K. Budniak$^{3,4}$,
  Z. Xue$^1$,
  H. De Libero$^5$,
  T. Thomson$^5$,\\
  M. Bosman$^{4,6}$, 
  G. Eda$^{3,7,8}$,
  H. Kurebayashi$^{1,9,10}$
  \& M. Cubukcu$^{1,2,\dagger}$
}

\date{
    \today
}

\begin{document}

\twocolumn[
  \maketitle             
  \begin{@twocolumnfalse} %

Antiferromagnetic (AFM) magnons in van der Waals (vdW) materials offer substantial potential for applications in magnonics and spintronics. In this study, we demonstrate ultra-strong magnon-magnon coupling in the GHz regime within a vdW AFM, achieving a maximum coupling rate of 0.91. Our investigation shows the tunability of coupling strength through temperature-dependent magnetic anisotropies. We compare coupling strength values derived from the gap size from the measured spectrum with those calculated directly through the coupling parameter and show that the gap size as a measure of coupling strength is limited for the ultra-strong coupling regime. Additionally, analytical calculations show the possibility to reach the deep-strong coupling regime by engineering the magnetic anisotropy. These findings highlight the potential of vdW AFMs as a model case to study magnetisation dynamics in low-symmetry magnetic materials.  


\vspace{16pt}
  \end{@twocolumnfalse}
    ]

\thispagestyle{fancy}
\setlength{\footskip}{15pt}

\section*{}

Within the expansive range of materials explored for spintronic applications, antiferromagnets (AFMs) have emerged as compelling candidates owing to their unique properties and potential to address challenges associated with traditional ferromagnetic (FM) devices \cite{jungwirth2016antiferromagnetic,baltz2018antiferromagnetic}. Characterised by antiparallel alignment of neighboring magnetic moments, net zero magnetisation and atomistic exchange coupling strength, AFMs offer inherent advantages such as insensitivity to external magnetic fields and ultra-fast spin dynamics \cite{gomonay2018antiferromagnetic}. However, the intricate nature of spin dynamics in AFMs presents both opportunities and challenges, underscoring the need for further investigation to fully harness their potential in spintronics.

Layered van der Waals (vdW) materials can be mechanically exfoliated down to monolayers relatively easily under controlled experimental conditions \cite{novoselov2004electric}. The recent discovery of long-range magnetic order at the two-dimensional (2D) limit, which does not follow the Mermin-Wagner theorem \cite{mermin1966absence} due to finite magnetic anisotropies, has stimulated an intense exploration of the growing variety of magnetic 2D materials \cite{gong2017discovery, huang2017layer}. Similar to synthetic antiferromagnets (SyAF) \cite{duine2018synthetic}, vdW AFM's such as CrCl$_3$ \cite{macneill2019gigahertz} and CrPS$_4$ \cite{peng2020magnetic} have a weak interlayer exchange coupling, which brings the typical terahertz (THz) dynamics of atomistic AFM's down to the gigahertz (GHz) regime. This characteristic makes them an ideal platform for studying a variety of spin dynamics phenomena using modern microwave techniques.

The interaction between magnons and other quasiparticles has garnered interest due to its fundamental importance in understanding complex quantum material interactions and its potential application in hybrid magnonics \cite{awschalom2021quantum,lachance2019hybrid}. Using the unique properties of magnons, these systems have shown significant promise in quantum information processing, storage, and sensing \cite{tabuchi2015coherent,lachance2020entanglement}. Due to the inherently weak dipolar interaction between magnons and photons, it is often difficult to achieve ultra-strong magnon-photon coupling without extensive engineering and optimisation \cite{golovchanskiy2021ultrastrong,niemczyk2010circuit}. The coupling rate, typically defined as the ratio of the gap size to the bare excitation frequency, is used to quantify the rate of information transfer between coupled modes and defines the interaction regimes. When the coupling rate increases above 0.1, the ultra-strong coupling regime is realised \cite{forn2019ultrastrong,Kockum2019,Diaz2019}. For light-matter systems in this regime, the rotating wave approximation does not hold and the gap size is no longer a good representation of the coupling strength. To account for this more accurate models are adopted, such as the quantum Rabi model and Hopfield model \cite{hopfield1958theory}. In ultra-strong coupling systems, the interaction Hamiltonian consists of both the co-rotating and counter-rotating terms. This permits exotic physical phenomena, such as nontrivial ground states and superradiant phase transitions \cite{wang2024ultrastrong,frisk2019ultrastrong,wang1973phase}. Recently the interaction between magnons themselves has emerged as a new avenue for investigating coupling phenomena \cite{macneill2019gigahertz, sud2023magnon,chen2018strong,dion2024ultrastrong}. Unlike magnon-photon coupling, a strong magnon-magnon coupling strength is in general expected due to their large spatial mode overlap \cite{wang2024ultrastrong}.

Ultra-strong magnon-magnon coupling has been experimentally observed in different material systems and explained by various symmetry breaking procedures from extrinsic origins (oblique dc fields) and intrinsic (Dzyaloshinskii-Moriya interaction and anisotropy) \cite{wang2024ultrastrong, sud2023magnon, liensberger2019exchange,makihara2021ultrastrong, wang2024ultrastrongDMI}. Using vdW AFMs, Li et al. \cite{ultrastrongmagnonCrPS4} demonstrated that the symmetry breaking due to magnetic anisotropy in \ce{CrPS4} enable ultra-strong magnon-magnon coupling that can be controlled by applying different  magnetic fields with respect to the crystalline axes. Moreover, the deep-strong coupling regime has so far not been achieved experimentally in magnon-magnon interactions, with a recent study in SyAF's reaching values near to this \cite{wang2024ultrastrong}.  

Quantifying the strength of magnon-magnon coupling remains a topic of considerable debate. The gap size is commonly used as a measure of magnon-magnon coupling strength, serving as a tool to define ultra-strong interactions in magnon-magnon systems \cite{wang2024ultrastrong, dion2024ultrastrong,liensberger2019exchange,makihara2021ultrastrong, ultrastrongmagnonCrPS4}. Recently, it has been shown that when intrinsic symmetry breaking induces the coupling, that an indirect gap opening occurs between the resonance modes with increased coupling strength than through extrinsic means \cite{li2021symmetry}. Additionally, a dc-field independent coupling strength parameter has been proposed for defining this coupling, which can be derived from material parameters in the case of SyAF with asymmetric magnetisation between the layers \cite{li2021symmetry}. 

In this work, we demonstrate ultra-strong magnon-magnon coupling in the vdW AFM material CrPS$_4$ in the GHz regime, achieving a coupling rate of 0.91. We show how the orthorhombic magnetic anisotropy lowers the symmetry of the magnetic system, thereby inducing magnon-magnon coupling between the two resonance modes. Furthermore, we investigate the temperature dependence of this coupling and report the magnetic parameters as a function of temperature. We demonstrate the tunability of coupling strength with temperature through controlling magnetic parameters, such as saturation magnetisation and the magnetic anisotropy. We compare coupling rate values derived from gap size with those calculated directly from the intrinsic coupling parameter and discuss the difference between the two. Our model predicts a viable strategy to reach the deep-strong coupling regime by changing magnetic anisotropies.

\begin{figure*}[!ht]
  \centering
  \includegraphics[keepaspectratio, width=1.1\linewidth]{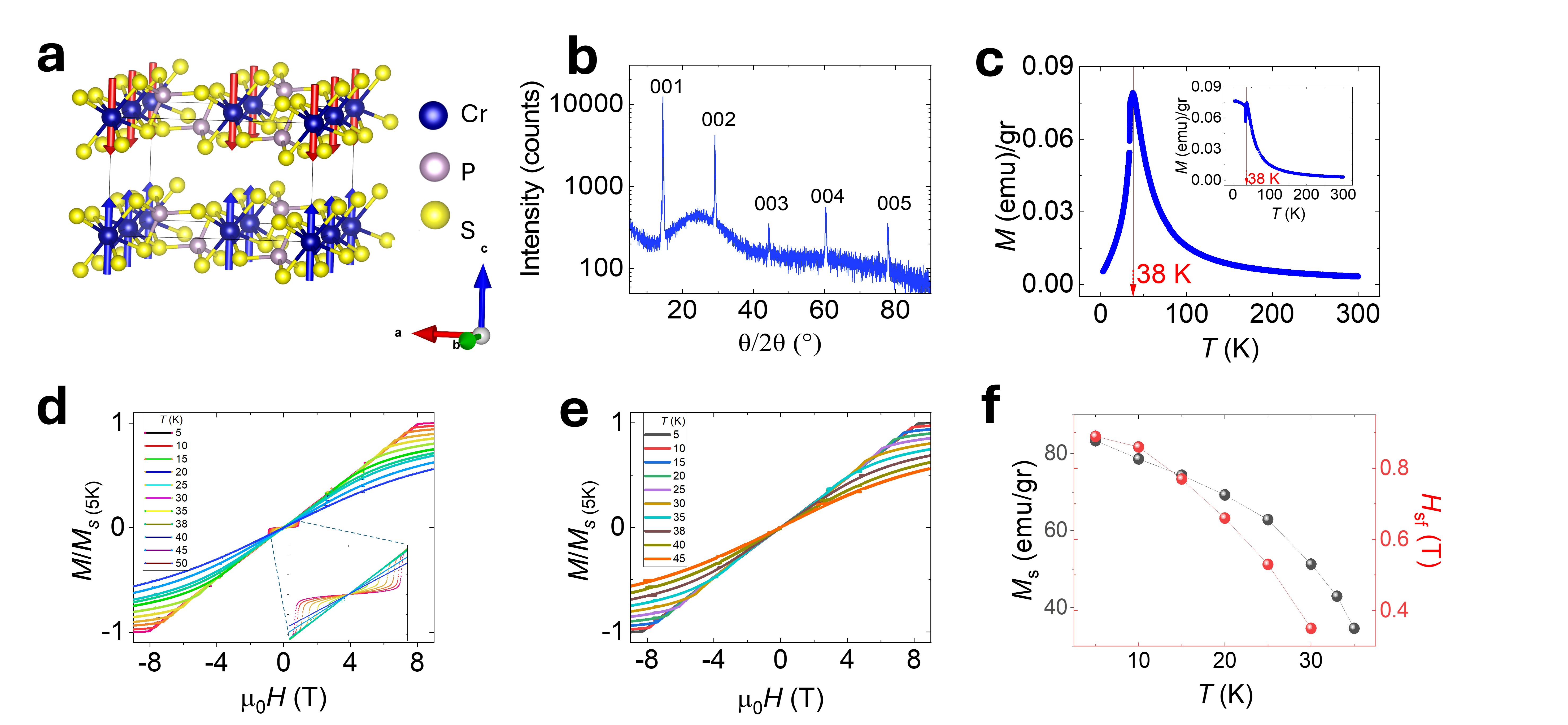}
  \caption{\textbf{Structure and magnetism in $\mathbf{CrPS}_4$}. a) Schematic illustration of crystal and magnetic structures of $\mathrm{CrPS}_4$. The blue and red arrows indicate the orientation of magnetic moments. b) X-ray diffraction of a single bulk CrPS$_{4}$ crystal. (c) Zero-field-cooled magnetisation ($M$) of a $\mathrm{CrPS}_4$ single crystal as a function of temperature ($T$), with a 10 mT magnetic field applied parallel to the crystallographic c-axis. The red line indicates $T_{\mathrm{N}}$. The inset illustrates the behavior under a magnetic field applied perpendicular to the c-axis. (d) Field-dependent normalised magnetisation $\left(M /\left(M_{S(5K)}\right)\right)$ at various temperatures with the magnetic field $\left(\mu_0 H\right)$ oriented parallel to the c-axis. The inset provides a zoomed-in image around spin-flop transition $\left(H_\mathrm{s f}\right)$. (e) Field-dependent normalised magnetisation at various temperatures with the field-oriented perpendicular to the $c$-axis. (f) Saturation magnetisation $\left(M_\mathrm{s}\right)$ and spin-flop transition field $\left(H_\mathrm{s f}\right)$ as a function of temperature ($T$).} 
  \label{Figure1}
\end{figure*}

\section*{RESULTS AND DISCUSSION}

\subsection*{Structure and Magnetism}

CrPS$_4$ is characterised by its semiconducting properties and an optically measured band gap of approximately 1.4 eV \cite{lee2017structural,louisy1978physical}. The crystal structure exhibits non-centrosymmetricity with monoclinic anisotropy, belonging to space group C$_2$, with crystal axes of lengths $a = 10.871$ Å, $b = 7.254$ Å, $c = 6.140$ Å, and $\beta = 91.88^\circ$ \cite{diehl1977crystal}. The crystal structure is illustrated in Fig. 1(a). In Fig. 1(b), we show X-ray diffraction of a single bulk CrPS$_{4}$ crystal. The position of 00x peaks is in agreement with work by Louisy et al. \cite{louisy1978physical} confirming the successful growth of high-quality CrPS$_{4}$.

We present the magnetometry results obtained from our bulk crystal in two field orientations, $H // c$ and $H // ab$, as functions of the applied magnetic field, $H$, and temperature, $T$. In Fig. 1(c), we depict the magnetisation, $M$, versus $T$ under a small applied field (10 mT) for $H // c$. A sharp peak at 38 K is observed followed by a steep decline, indicative of the N\'{e}el transition temperature, $T_\mathrm{N}$, consistent with that in the literature \cite{peng2020magnetic}. Figures 1(d) and 1(e) plot $M$ versus $H$ for the $H // c$ and $H // ab$, respectively, at $T$ ranging from 5 K to 50 K. For $T$ above 10 K, a prominent non-saturating background is observed. From these, we extract the saturation magnetisation as a function of $T$. In the $H // c$ orientation, a characteristic spin-flop transition is evident due to the out-of-plane magnetic easy axis of the material \cite{peng2020magnetic}. The spin-flop transition is indicated by the plateau in the moment until reaching a critical field, $H_\mathrm{sf}$. Beyond $H_\mathrm{sf}$, a sharp increase in $M$ is observed as the sublattices become canted towards the direction $H$. With increasing $H$, a linear rise in $M$ is measured until saturation, as the angle between the sublattices and $H$ decreases, ultimately saturating in the forced FM phase around 8 T at 5 K. The extracted values of $M_\mathrm{s}$ and $H_\mathrm{sf}$ are plotted as functions of temperature in Fig. 1(f), where both values decrease with increasing temperature.

\subsection*{Tunable Ultra-Strong Magnon–Magnon Coupling Approaching the Deep-Strong Regime}

We focus on ferromagnetic resonance (FMR) measurements conducted below the $T_\mathrm{N}$ for two orientations: $H \perp c$ with $\theta \approx 90^\circ$ and $45^\circ$, where $\theta$ is the angle between the $b$ axis and $H$. The crystal axis of the bulk crystal is determined visually, considering the straight edge of the crystal as the crystal growth aligns with the preferred $b$ axis \cite{peng2020magnetic}. This method of determination differs from the preferential edges observed during exfoliation, which typically exhibit an angle of $33.75^\circ$ away from the $a$ axis \cite{lee2017structural, joe2019dominant}.

\begin{figure*}[!ht]
  \centering
 
  \includegraphics[keepaspectratio,width=1\linewidth]{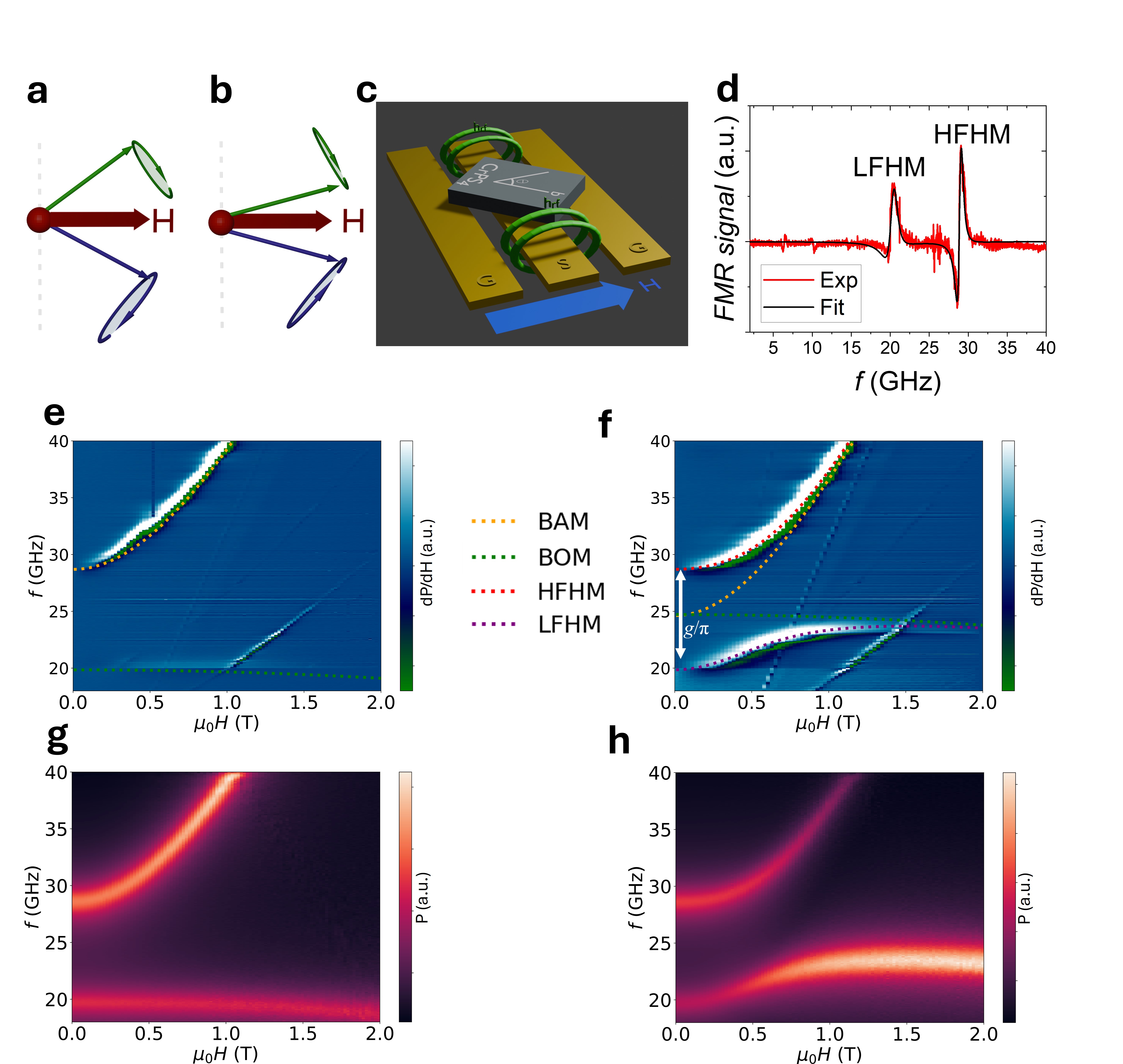}
  \caption{\textbf{Spin dynamics in $\mathbf{CrPS}_4$ under an in-plane magnetic field}. (a-b) Schematic illustration of the magnetic moments' precession in the acoustic mode (in-phase precession) (a) and the optical mode (out-of-phase precession) (b). (c) Schematic of the configuration between coplanar waveguide (CPW), $\mathrm{CrPS}_4$ crystal, and external in-plane magnetic field ($H$). The angle between $H$ and the $b$ axis is shown by $\theta$. (d) Typical ferromagnetic resonance (FMR) spectra (red line) obtained at 15 K and 200 mT and their derivated Lorentzian fitting (black line) by using  Eq. S9 (see SI). (e-f) 2D maps of the FMR spectra obtained at 15 K, with the magnetic field applied ($\mu_0 H$) at $\theta=90^{\circ}$  (e) and $\theta=45^{\circ}$ (f). The fitted resonance modes are denoted as BAM, BOM, HFHM, and LFHM, representing the bare acoustic mode, bare optical mode, high-frequency hybrid mode, and low-frequency hybrid mode, respectively. The gap size $g/\pi$ is depicted in (f) at 0 T where the uncoupled modes would intersect. (g-h) Simulated FMR spectra obtained with the parameters extracted from experimental data with $\mu_0 H$ at $\theta=90^{\circ}$ (g) and $\theta=45^{\circ}$ (h).}
  
\end{figure*}

\begin{figure*}[!ht] 
  \centering
  \includegraphics[width=1\textwidth]{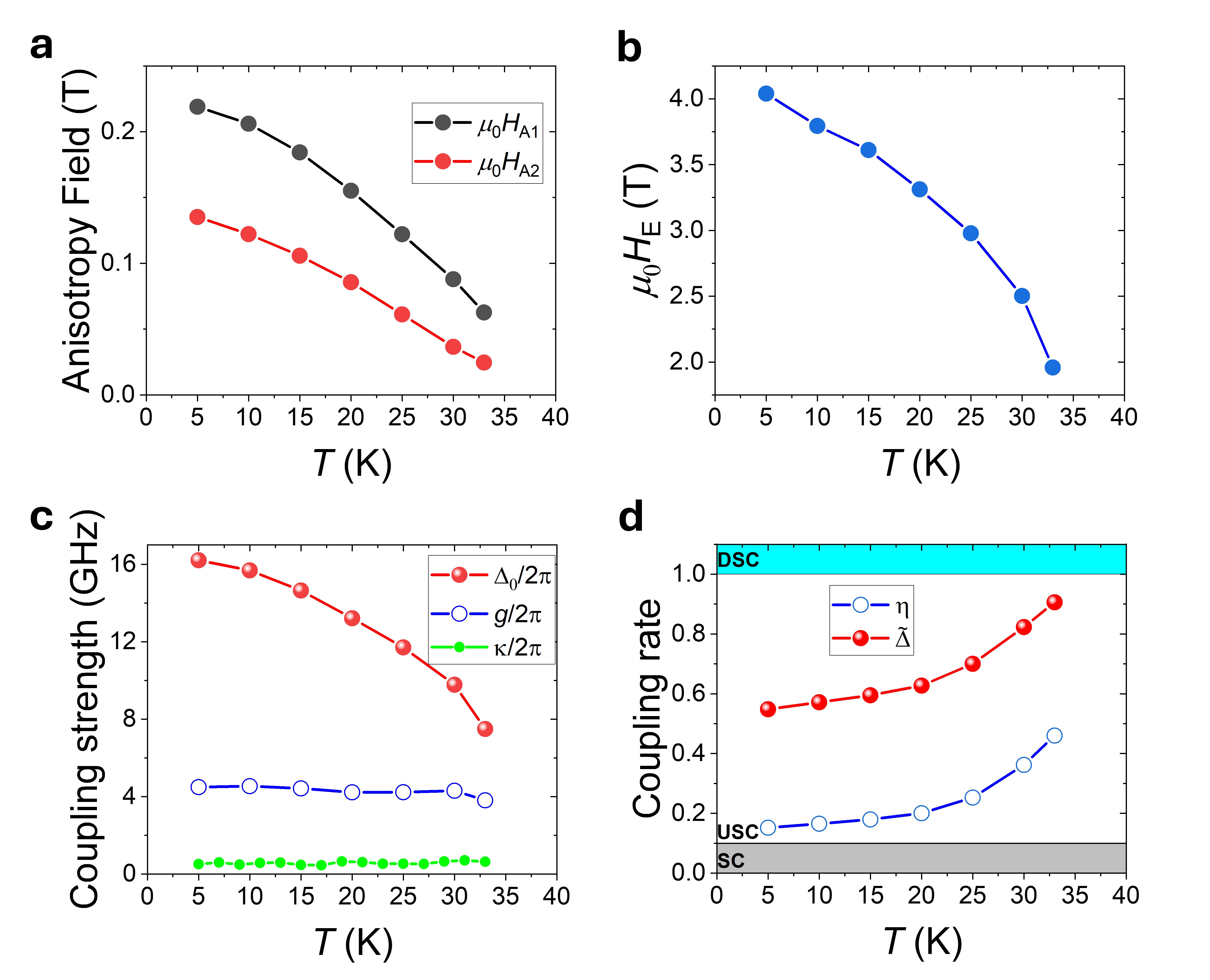} 
  \caption{\textbf{Temperature dependence of magnetic parameters and coupling} (a-b) Plot showing the temperature dependence of $\mu_0 H_{A1}$, $\mu_0 H_{A2}$ (a) and $\mu_0 H_{E}$ (b). (c) Temperature dependence of the intrinsic coupling strength, $\Delta_0/2\pi$, coupling strength, $g/2\pi$, and the line width, $\kappa/2\pi$. A significant increase between the two methods of calculating coupling strength is observed. (d) Temperature dependence of the coupling rate determined through the two methods, gap size $\eta$ and intrinsic coupling $\Tilde{\Delta}$. The intrinsic coupling rate is seen to approach the deep-strong coupling regime with temperature. SC, USC, and DSC represent strong coupling, ultra-strong coupling, and deep-strong coupling regimes, respectively.
}
  \label{fig:3}
\end{figure*}

We now present a detailed study of the temperature dependence of the resonances measured in the in-plane orientation with respect to the crystal axis, $b$, at angles $\theta \approx 90^\circ$ and $45^\circ$. Measurements in the out-of-plane orientation are included in the Supplementary Infomation (SI) (Figs. S1 and S2).

To model the spin dynamics, we employ the coupled Landau-Liftshitz-Gilbert (LLG) equation for two macrospin sublattices coupled through an interlayer exchange interaction, with zero damping for simplicity. Starting with Keffer's free energy equation for orthorhombic anisotropy and neglecting the terms dependent upon the relative positions of the sublattice moments \cite{keffer1952theory}:

\begin{equation}
    F_\mathrm{A} = K_1/2(\alpha_1^2 + \beta_1^2) + K_2/2(\alpha_2^2 + \beta_2^2)
\end{equation}

where $F_\mathrm{A}$ is the anisotropy component free energy, $K_1$ and $K_2$ are the anisotropy constants related to the $a$ and $b$ axis, $\alpha_1$, $\alpha_2$, $\beta_1$ and $\beta_2$ are the cosines of the unit magnetisation for the two sublattices. The LLG equation can then be written as follows: 

\begin{equation}
\begin{aligned}
& \frac{d \hat{m}_\mathrm{A}}{d t}=-\mu_0 \gamma \hat{m}_\mathrm{A} \times(\mathbf{H}-H_\mathrm{E} \hat{m}_\mathrm{B}-M_\mathrm{s}\left(\hat{m}_\mathrm{A} \cdot \hat{z}\right) \hat{z} \\
&\left.-\mathrm{H}_{\mathrm{A} 1}\left(\hat{\mathrm{m}}_A \cdot \hat{\mathrm{x}}\right) \hat{\mathrm{x}}-\mathrm{H}_{\mathrm{A} 2}\left(\hat{\mathrm{m}}_\mathrm{A} \cdot \hat{y}\right) \hat{y}\right), \\
& \frac{d \hat{m}_\mathrm{B}}{d t}=-\mu_0 \gamma \hat{m}_\mathrm{B} \times(\mathbf{H}-H_\mathrm{E} \hat{m}_\mathrm{A}-M_\mathrm{s}\left(\hat{m}_\mathrm{B} \cdot \hat{z}\right) \hat{z} \\
&\left.-\mathrm{H}_{\mathrm{A} 1}\left(\hat{\mathrm{m}}_\mathrm{B} \cdot \hat{\mathrm{x}}\right) \hat{\mathrm{x}}-\mathrm{H}_{\mathrm{A} 2}\left(\hat{\mathrm{m}}_\mathrm{B} \cdot \hat{y}\right) \hat{y}\right))
\end{aligned}
\end{equation}

where ${\mathrm{m}}_\mathrm{A}$ and ${\mathrm{m}}_\mathrm{B}$ are the moments of the two sublattices, $\gamma$ is the gyromagnetic ratio and $H_\mathrm{E}$ denotes the interlayer exchange field. The anisotropy fields, $H_\mathrm{A1}$ and $H_\mathrm{A2}$, are defined as $K_1 / \mu_0 M_\mathrm{s}$ and $K_2 / \mu_0 M_\mathrm{s}$, respectively. In this model, the crystal axes $a$, $b$, and $c$ are aligned with the coordinate axes $x$, $y$, and $z$, respectively. The unit vectors of this coordinate system are denoted by $\hat{x}$, $\hat{y}$, and $\hat{z}$.

After Fourier transformation, the LLG equation can be treated as an 2$\times$2 eigenvalue problem with the characteristic equation of:

\begin{equation}
\begin{aligned}
\label{eq: eigenvalue}
\begin{vmatrix}
\omega_\mathrm{BAM}^2 (H, \theta) - \omega^2 & \Delta^2 (H, \theta) \\
\Delta^2 (H, \theta) & \omega_\mathrm{BOM}^2 (H, \theta) - \omega^2
\end{vmatrix} 
 = 0
\end{aligned}
\end{equation}

where $\Delta$ denotes the intrinsic magnon-magnon coupling term and $\omega_{\text{BAM}}$ and $\omega_{\text{BOM}}$ are the bare acoustic and optical mode frequencies respectively. These precession modes are depicted in Fig. 2(a) and 2(b) respectively.

Solutions to this reveal that when $\theta \neq  0^\circ$ an anti-crossing gap opens up around the intersection between the two modes through mode hybridisation. The anti-crossing gap is the signature of such hybridisation and the gap size between the hybridised modes is defined here as $g/\pi$ which we take at zero field as this is the point where the uncoupled modes would intersect at $\theta = 45^\circ$ . Equation 3 can then be used to fit experimental data as described in the following section. 
Solving Eq. 3 for $\theta = 0^\circ$, we can obtain the frequency versus field dependence of the bare acoustic and optical modes, BAM and BOM respectively, given by Eq. S8 in the SI.

The experimental setup is depicted in Fig. 2(c). A typical frequency swept spectra is shown in Fig. 2(d) which can be fitted with two differentiated Lorentzian peaks (see SI, Eq. S9), from which the resonance frequency can be extracted.  
In Fig. 2(e), we plot a 2D map of the spectra obtained when $\theta \approx 90^\circ$ at 15 K. We observe two modes that can be fitted by BAM and BOM described in Eq. 3. Furthermore, simulations at $\theta = 0^\circ$ shown in Fig. S3(b) in the SI demonstrate the intersection of the two modes along this high symmetry axis. This observation confirms the absence of a symmetry-breaking condition in these orientations.

In Fig.~2(f), we plot the 2D map of the spectra obtained at 15 K in the in-plane orientation with $\theta \approx 45^{\circ}$. Here, we clearly observe magnon-magnon coupling, as evidenced by the opening of the anticrossing gap, which is comparable to the bare magnon frequencies and hence in the ultra-strong coupling regime. These two hybridised modes can be well-fitted with the computation of Eq. 3. 
\newpage
Two other modes are also observed in the 2D map. The mode that crosses the LFHM at approximately 1.5~T in Fig.~2(f) is identified as a harmonic resonance of the HFHM mode and appears at half the resonant frequency of the HFHM. Additionally, double, quarter, and eighth harmonics are observed, which are fully depicted in Fig.~S4 of the SI. The other mode, a linear mode reaching 40~GHz at approximately 1.2~T, is explained by background absorption originating from the waveguide itself. Reference measurements are provided to confirm this (see SI, Fig.~S5).

In Figs.~2(g)-(h), we present micromagnetic simulations carried out in Mumax3, for $\theta = 90^\circ$ and $\theta = 45^\circ$, respectively. The simulation is carried out using the magnetic parameters determined from fitting at
$15$ K and shows good agreement with the experimental data as shown in Fig. 2(e-f).

By fitting the experimental resonant frequencies for the coupled modes, we can extract $H_\mathrm{A1}$ and $H_\mathrm{A2}$. Furthermore, we calculate the coupling strength as half the gap size, $g/2\pi$, of the coupled modes at the field at which the two uncoupled modes would intersect (0 T when $\theta = 45^\circ$).

The values obtained for $H_\mathrm{A1}$ and $H_\mathrm{A2}$, are presented in Fig. 3(a). Throughout all temperatures, the value of $H_\mathrm{A1}$ remains larger than $H_\mathrm{A2}$. Both anisotropy fields follow a decreasing function with temperature as is typical of magneto-crystalline anisotropies \cite{callen1966present}. In Fig. 3(b), we plot the values of $H_E$ versus $T$ which also decreases as a function of temperature as would be expected and agreeing with recent results using magnetic torque \cite{seo2024probing}.

It is typical to use the coupling rate, $\eta$, by comparing the anticrossing gap size to the uncoupled excitation frequency, where $\eta$ is given by the following: 

\begin{equation}
    \eta = \frac{g}{2\pi f_\mathrm{r}}
\end{equation}

Here, $f_\mathrm{r}$ represents the bare excitation frequency at the crossing of the uncoupled BAM and BOM, for $\theta = 45^\circ$ this occurs at 0 T.

Similarly, we can now define the intrinsic coupling rate, $\Tilde{\Delta}$, as follows from the intrinsic magnon-magnon coupling strength defined in Eq. 3:

\begin{equation}
    \Tilde{\Delta} = \frac{\Delta_0}{2 \pi f_\mathrm{r}}
\end{equation}
where $\Delta_0$ is the value of $\Delta$ at the point of intersection of the uncoupled modes.

In Fig. 3(c), we compare the values of $\Delta_0/2\pi$ and $g/2\pi$ to the linewidth, $\kappa/2\pi$, of the hybridised modes. The linewidth is taken at a field of 200 mT. No clear field or temperature dependence of the linewidth is observed due to large extrinsic broadening (see SI, Fig. S6) as is common in bulk vdW samples where inhomogeneity across the sample makes linewidth analysis difficult \cite{xu2023electrical}. In both $\Delta_0/2\pi$ and $g/2\pi$, we show that the coupling strengths are significantly larger than  $\kappa/2\pi$  meeting the criteria for strong coupling \cite{frisk2019ultrastrong,Diaz2019}. Furthermore, we observe that  $\Delta_0/2\pi$ is significantly larger than  $g/2\pi$ over the the entire temperature range. This agrees with the findings of Li et al.~\cite{li2021symmetry} in asymmetric SyAFs, which suggest that the gap size should not be used to quantify the coupling strength in cases of intrinsic symmetry-breaking-induced magnon-magnon coupling.

\begin{figure*}[!ht]
  \centering
  \includegraphics[width=1\linewidth]{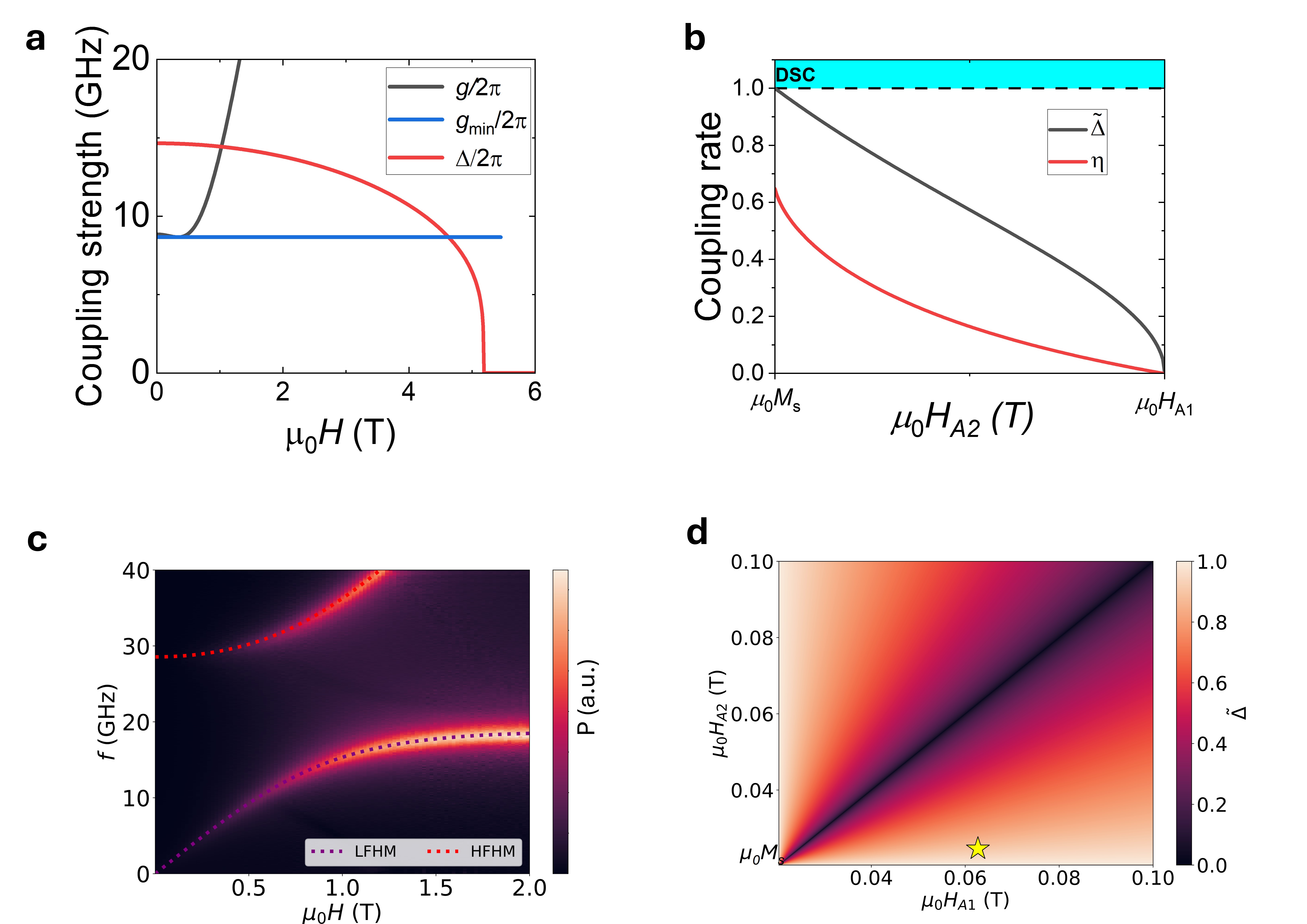} 
  \caption{\textbf{Tunable coupling rate via magnetic anisotropy up to deep-strong coupling.} (a) Field dependence of ${\Delta}/2\pi$ compared to the $g/2\pi$ at 15K, where $g_{min}/2\pi$ is the half of the minimum gap size. (b) Analytical values of $\eta$  and $\Tilde{\Delta}$ as a function of $H_{A2}$, when $\mu_0 H_{A2}$ is equal to $\mu_0 M_s$ the deep strong regime is reached. (c) Micromagnetic simulation of the FMR spectra for the deep-strong coupling case, which shows good agreement with the analytical solutions. (d) 2D plot of $\Tilde{\Delta}$ for a fixed value of $M_\mathrm{s}$ at 33 K as a function of $H_\mathrm{A2}$ and $H_\mathrm{A1}$, demonstrating the potential tunability of the coupling rate. The star indicates the position and value of $\Tilde{\Delta}$ observed experimentally at 33 K.}
  \label{fig:4}
\end{figure*}

We now compare the coupling rates as a function of temperature obtained through the two methods, $\eta$ and $\Tilde{\Delta}$, as shown in Fig. 3(d). We observe a significant difference between the two methods at all temperatures, with $\Tilde{\Delta}$ being more than twice the value of $\eta$. Furthermore, $\Tilde{\Delta}$ increases as a function of $T$, from a minimum of 0.55 at 5 K to a maximum of 0.91 at 33 K, approaching the deep-strong regime as the temperature approaches $T_N$. This temperature dependence arises from the interplay between the magnetic parameters as a function of temperature, as we demonstrate below.

In Fig.~4(a), we plot the field dependence of ${\Delta} / 2\pi$ compared to $g / 2\pi$ as a function of field at an example temperature of 15 K. We demonstrate a notable difference between the two parameters at the intersection field (0 T when $\theta = 45^\circ$). Additionally, we observe that ${\Delta} / 2\pi$ is field-dependent, as it is determined by the canting angle between the sublattice moments. ${\Delta} / 2\pi$ reaches a maximum when the moments are antiparallel at zero field and becomes zero when the canting angle exceeds $45^\circ$. This contrasts with the field-independent ${\Delta} / 2\pi$ introduced for the SyAF, where the intrinsic symmetry breaking arises from differing $M_\mathrm{s}$ values between the layers and has no field dependence~\cite{li2021symmetry}.

To further investigate the coupling rate, we compare the simulated coupling rate obtained from both methods as a function of $H_{A2}$, as shown in Fig.~4(b). Here, $\eta$ is observed to increase as $H_{A2}$ decreases, reaching a value of approximately 0.65. At this value, $H_{A2}$ becomes less than $M_s$, and micromagnetic simulations show that the system transitions into an in-plane easy-axis system (see SI, Fig.~S7). On the other hand, $\Tilde{\Delta}$ increases, reaching a maximum value of 1 when $H_{A2}$ equals $M_\mathbf{s}$. We define this point as the critical coupling strength, corresponding to the deep-strong regime, similar to observations in anisotropic SyAFs~\cite{wang2024ultrastrong, Kockum2019, casanova2010deep}.  

In Fig.~4(c), we present the micromagnetic simulation of this deep-strong coupling case. In this simulation, the lower hybridized mode is observed to reach 0 GHz at 0 T, indicating the maximum achievable coupling rate. These simulated modes are well fitted by analytical solutions, setting $\mu_0 H_{A2}$ to $\mu_0 M_s$, which yields a coupling rate of 1.

We now derive the analytical solution for $\Tilde{\Delta}$ from the Eq. 3 and Eq. S7 in the SI, yielding the following expression: 

\begin{strip}
    
\begin{align}
\Tilde{\Delta} = \frac{\sqrt{(H_{A1} - H_{A2}) H_E}}{\sqrt{H_{A2} (H_E - M_s) + H_{A1} (H_{A2} + H_E - M_s) + M_s (-2 H_{E} + M_s)}} 
\end{align}
\end{strip}

Eq. 6, provides insight into how to precisely tune the coupling rate through control of the two anisotropy values. In Fig.~4(d), we present a 2D plot of the $\Tilde{\Delta}$ as a function of $\mu_0 H_{A1}$ and $\mu_0 H_{A2}$ for the values of $\mu_0 M_\mathrm{s}$ and $\mu_0H_\mathrm{E}$ at 33 K. This demonstrates the possibility to tune the coupling strength from uncoupled to critically coupled regime through manipulation of the two anisotropy terms. Avenues for achieving this precise tuning include ionic gating, intercalation, or strain, all of which have been shown to have considerable effects on anisotropy in vdW magnetic materials \cite{verzhbitskiy2020controlling,khan2024spin,tang2020tunable,webster2018strain,qi2023giant}. 

\section*{CONCLUSIONS}

In conclusion, our investigation has revealed the temperature dependence of the ultra-strong magnon-magnon coupling in the vdW AFM CrPS$_4$, approaching the deep-strong regime with a maximum coupling rate of 0.91 achieved at 33 K. This coupling is attributed to the interplay of orthorhombic anisotropies, saturation magnetisation, and the exchange field. We evaluated the coupling rate using two approaches: the gap size $\eta$ and the intrinsic coupling rate $\Tilde{\Delta}$. Our findings indicate that, in systems with intrinsic symmetry-breaking anisotropy, these two methods yield notably different coupling rates.
We further provide an analytical description of $\Tilde{\Delta}$ in the case of orthorhombic anisotropy, demonstrating how tuning anisotropy values to a critical point could allow for realisation of deep-strong coupling. This study not only enhances our understanding of the spin dynamics in vdW AFMs, but also demonstrates the potential for vdW AFMs as an ideal system to experimentally reach deep-strong magnon-magnon coupling.

\bibliography{main.bib}
\section*{METHODS}

\subsection*{Material Growth and Sample Characterisation}
Chromium thiophosphate (CrPS$_4$)  was synthesized in a quartz ampoule via physical vapor transport (PVT) – without any transporting agent \cite{budniak2020exfoliated}. About half of a gram of stochiometric elements mixture (metal chromium powder, red phosphorus powder, and elemental sulfur powder (all purchased from Sigma-Aldrich), Cr:P:S = 1:1:4) was ground in agate mortar, moved into quartz ampoule, evacuated to high vacuum (below $5 \times 10^{-6}$ mbar) by turbomolecular pump and closed by a flamer. The sealed ampoule was put into a two-zone furnace, that was calibrated in a way that the mixture of elements was kept at 750 °C and the deposition zone was 710 °C. The ampoule was warmed from room temperature to designated ones within five hours, after four days the furnace was turned off and the sample was allowed to cool down naturally. Then the ampoule was opened, and only recrystallized, large, pure CrPS${_4}$ crystals from the deposition zone (710 °C) were collected. 

The diffractogram of a single, as-grown CrPS$_4$ crystal placed in the preferential (001) orientation was recorded using a thin-film X-ray diffractometer (TL-XRD), Bruker D8 Advance, equipped with a Cu anode and a parallel beam.  

Energy-dispersive X-ray spectroscopy (EDS) and electron micrographs were obtained using a scanning electron microscope (SEM), Zeiss Sigma 300, equipped with an Oxford EDS AztecLive 60 mm\(^2\) area EDS detector. The experiments were performed at an acceleration voltage of 20 kV without any sample coating.

\subsection*{Measurement Technique}

To investigate the magnetism and spin dynamics in the vdW AFM CrPS$_{4}$, we performed magnetometry and ferromagnetic resonance (FMR) measurements using a Quantum Design Physical Property Measurement System (PPMS). This PPMS configuration allows for the application of magnetic fields up to 9 T and provides precise temperature control down to 1.8 K. The system is equipped with a vibrating sample magnetometer (VSM) and a NanOsc CryoFMR probe for coplanar waveguide (CPW)-FMR spectroscopy, covering a frequency range of 2–40 GHz.  

Broadband microwave spectrum measurements were performed at low temperatures using a CPW with a center conductor width of $w = 2$ mm. We conducted field-modulated, frequency-sweep measurements at a constant magnetic field and temperature to generate 2D dispersion maps.

\subsection*{Micromagnetic Simulation}
 Dynamic micromagnetic simulations were performed using the Mumax3 package \cite{vansteenkiste2014design}. CrPS$_{4}$ was modelled on a $5\times5\times25$ nm$^{3}$ cell grid with dimensions $150\times250\times50$ nm$^{3}$. The upper and lower cells in the out of plane direction were antiferromagnetically coupled with an exchange scaling of -3.04 and exchange stiffness of $10\times10^{-12}$ J/m. Gilbert damping was set to $0.01$ and the simulation temperature was $0$ K.
The simulation is carried out using the magnetic parameters determined from magnetometry. FMR spectra are obtained using sinc field pulses in combination with an applied external dc field. The sinc pulse period was set to 10 ns and the time-varying magnetisation responses were Fourier transformed to obtain the final spectra. 

\subsection*{Data Availability} 
All relevant data are available from the corresponding author upon reasonable request.

\subsection*{ACKNOWLEDGEMENTS}
C.W.F.F. and M.C. acknowledge support from the UK National Quantum Technologies and National Measurement System programmes. C.W.F.F. and H.Y. thank EPSRC for support through EPSRC DTP Case studentship (EP/T517793/1). H.K. acknowledges support from EPSRC via EP/T006749/1 and EP/V035630/1. A.K.B., M.B. and G.E. acknowledge support from the Ministry of Education (MOE), Singapore, under AcRF Tier 3 (MOE2018-T3-1-005). We thank O. Lee for assistance with Blender.

\subsection*{Author Contributions}
C.W.W.F., H.K., and M.C. conceived and planned the project. C.W.W.F. and M.C. conducted VSM, FMR, as well as data analysis. H.Y. and C.W.W.F. carried out MuMax3 simulations. A.K.B. grew the CrPS$_4$ crystal under the supervision of G.E. and performed XRD and SEM with EDS measurements under supervision of M.B.. C.W.W.F. and M.C. wrote the manuscript. All authors contributed to discussions on the results and provided feedback on the manuscript.

\subsection*{Competing Interests}
The authors declare no competing interests.

\section*{Additional information}

\textbf{Supplementary information}. 
The online version contains supplementary material available at (link).

\raggedright
\textbf{Correspondence} and requests for materials should be addressed to M.C..

\end{document}